\documentclass[showpacs,amsmath]{revtex4}
\usepackage{graphicx}%
\usepackage{dcolumn}
\usepackage{amsmath}
\usepackage{here}
\begin{document}
%\DeclareRobustCommand{\baselinestretch{2}}

\title{Spatial vector solitons in nonlinear photonic crystal fibers}

\author{Jos\'e R. Salgueiro and Yuri S. Kivshar}

\affiliation{Nonlinear Physics Centre and Center for Ultra-high
bandwidth Devices for Optical Systems (CUDOS), Research School of
Physical Sciences and Engineering, The Australian National
University, Canberra ACT 0200, Australia}

\author{Dmitry E. Pelinovsky}

\affiliation{Department of Mathematics, McMaster University,
Hamilton, Ontario, Canada L8S 4K1 }

\author{Ver\'onica Sim\'on and Humberto Michinel}

\affiliation{\'Area de \'Optica, Facultade de Ciencias de Ourense.
As Lagoas s/n, 32004 Ourense, Spain}

\begin{abstract}
We study {\em spatial vector solitons} in a photonic crystal fiber
(PCF) made of a material with the focusing Kerr nonlinearity. We
show that such two-component localized nonlinear waves consist of
two mutually trapped components confined by the PCF linear and the
self-induced nonlinear refractive indices, and they bifurcate from
the corresponding scalar solitons. We demonstrate that, in a sharp
contrast with an entirely homogeneous nonlinear Kerr medium where
both scalar and vector spatial solitons are unstable and may
collapse, the periodic structure of PCF can stabilize the otherwise
unstable two-dimensional spatial optical solitons. We apply the matrix
criterion for stability of these two-parameter solitons, and verify
it by direct numerical simulations.
\end{abstract}


\maketitle

\section{Introduction}

Photonic crystal fibers (PCF) have attracted much interest due to
their intriguing properties, many potential applications, as well as
the recent development of successful technologies for their
fabrication with engineered linear and nonlinear
properties~\cite{Russell2003,Knight}. Photonic crystal fibers are
characterized by a conventional cylindric geometry with a
two-dimensional lattice of air-holes running parallel the fiber
optical axis. Such PCF structures share the propagation properties
of {\em photonic crystals}, based on the existence of the frequency
gap with the transmission suppressed due to the Bragg scattering, as
well as the properties of conventional {\em optical fibers}, due to
the presence of a defect in the structure acting as a PCF core. Some
of the PCF intriguing characteristics include the possibility to
design single-moded PCFs independently on the light frequency even
for a large core, allowing the guidance of high powers what makes
PCFs very suitable for amplifiers or laser cavity applications. On
the other hand, there exists an upper cutoff frequency by means of a
reduction of the core index, and this also allows a very flexible
control on the dispersion properties, supporting large shifts of the
zero-dispersion point, and birefringence, which can be made much
higher than in conventional fibers by a proper design.

In PCFs, light confinement is restricted to the core of the fiber
and therefore nonlinear effects, such as light self-trapping and
localization in the form of spatial optical solitons~\cite{book}, 
may become important. The stabilizing effect of periodic media for
optical solitons has been observed in a number of cases. In 
particular, one-dimensional vector solitons that are unstable in
uniform media were made stable in a medium with a periodic
modulation of the refractive index~\cite{Kartashov2004}. Also,
discrete vector solitons where experimentally observed in
two-dimensional optically-induced photonic
lattices~\cite{Chen2004}. Similar to the case of two-dimensional
nonlinear photonic crystals~\cite{Mingaleev2001}, it has been
recently demonstrated numerically that a PCF can support and
stabilize both fundamental and vortex spatial optical
solitons~\cite{Ferrando2003,Ferrando2004}. In a sharp contrast
with an entirely homogeneous nonlinear Kerr medium where spatial
solitons are unstable and may collapse, it was shown that the
periodic structure of PCF can stabilize the otherwise unstable
two-dimensional spatial optical solitons.

In this paper, we make a further step forward in the study of
nonlinear effects in PCFs, in comparison with the recent
analysis~\cite{Ferrando2003,Ferrando2004}, and analyze the existence
and stability of spatial vector solitons in PCFs. In general, vector
solitons are defined as two-component mutually trapped localized
beams whose properties may differ substantially from the properties
of one-component scalar solitons~\cite{book}. In addition,
two-dimensional vector solitons are known to be unstable in the
nonlinear Kerr medium (see, e.g., Ref.~\cite{Malmberg2000}). In
contrast, as we show in this paper, the periodic modulation of the
refractive index in PCF provides an effective physical mechanism to
stabilize the otherwise unstable two-dimensional spatial optical
solitons. We study the stability of these two-parameter solitons and
apply the matrix stability criterion that is then verified by direct
numerical simulations.

The structure of this paper is the following. First, In Sec.~II we
introduce our physical model that is characterized by an effective
potential created by the PCF environment and also describe the
nonlinear interaction between the beam components. Then, in Sec.~III 
we introduce our numerical method to find the classes of
spatially localized modes existing in the nonlinear core of the PCF.
In Section~IV we describe the family of two-component spatial
solitons. Finally, in Sec.~V the stability of both one- and 
two-component solitons is analyzed.

\section{Model}

We consider a simple model of PCF that describes, at a given
frequency, the spatial distribution of light in a nonlinear
dielectric material with a triangular lattice of air holes in a
circular geometry. We assume that the PCF material possesses a
nonlinear Kerr response, and the hole at the center is filled by the
same material creating a nonlinear defect, as shown in
Fig.~\ref{scalar}(a,b). In the substrate material of the fiber, the
linear refractive index is $n_s$, whereas inside the holes it is
$n_a$. Air holes have radius $r$. We consider the case when the PCF
core guides two modes or two orthogonal polarizations. In the
nonlinear regime, the mutual interaction between these two modes is
described by the system of coupled equations,
\begin{equation}
\label{model_equations}
   \begin{array}{l} {\displaystyle
     i \frac{\partial \psi_{1}}{\partial z} + \Delta \psi_1 +
     n_a \psi_1 + V(x,y) (\delta + |\psi_{1}|^2+\mu |\psi_{2}|^2)\psi_1 =0,}\\*[9pt]
        {\displaystyle
    i \frac{\partial \psi_{2}}{\partial z} +\Delta \psi_2 +
     n_a \psi_2 + V(x,y) (\delta + |\psi_{2}|^2+\mu |\psi_{1}|^2)\psi_2 =0,}
   \end{array}
\end{equation}
where $\psi_1$ and $\psi_2$ are two components (or two
polarizations) of the electric field, $\Delta$ is a transversal
Laplace operator in $(x,y)$, $\delta=n_s-n_a$, and $V(x,y)$ is an
effective potential describing the defect and the
lattice of holes in the transverse plane $(x,y)$. We normalize
$V = 1$ in the material outside the holes, and $V = 0$ in the holes.
The nonlinear incoherent interaction between the components is
described by the parameter $\mu$.

To find stationary two-dimensional nonlinear modes of PCF, we look
for the solutions in the form
$$
\psi_1(x,y,z)=u(x,y)e^{i\beta z}, \; \psi_2(x,y,z)=v(x,y)
e^{i\gamma z},
$$
and obtain the following coupled system of $z$-independent
differential equations:
\begin{equation}
\label{model_stationary}
   \begin{array}{l} {\displaystyle
      \beta u =\Delta u + n_a u + V(x,y) \left(\delta + u^2 + 
		\mu v^2 \right) u,}\\*[9pt]
        {\displaystyle
      \gamma v =\Delta v + n_a v + V(x,y) \left(\delta + v^2 + 
		\mu u^2 \right) v.}
   \end{array}
\end{equation}
The model (\ref{model_stationary}) describes the stationary
distribution of a two-component field in an inhomogeneous
nonlinear medium, in a planar geometry. Without the external
potential, the vector solitons in both one- and two-dimensional
cases have been studied earlier~\cite{book}. However, the lattice
of air-holes and the central defect break the radial symmetry of
the problem, and the corresponding vector solitons are not
radially symmetric.

\section{Numerical method}

In order to find the solutions for nonlinear localized modes, we
consider a rectangular domain of the $(x,y)$ and apply a
finite-difference scheme, taking $n$ and $m$ uniformly
distributed samples of the variables
$x$ and $y$, respectively, in order to cover all the domain.
Denoting those samples as $x_i, 1\leq i\leq n$, and $y_j,
1\leq j\leq m$, at each mapped point $(x_i,y_j)$ of the domain we
consider the corresponding samples for all the functions defined in
the equations: $u_{ij}=u(x_i,y_j)$ and, similarly, the second
component $v_{ij}$, and the potential $V_{ij}$. Substituting these
re-defined variables into the model~(\ref{model_stationary}), and
imposing homogeneous boundary conditions in all four edges of the
domain, we obtain an algebraic nonlinear problem of $2\times
n\times m$ equations with the same number of unknowns $u_{ij}$ and
$v_{ij}$.

In order to make the notation more compact, the samples
corresponding to different functions, which constitute $n\times m$
matrices, are rearranged concatenating the columns of the matrices
to produce big column vectors of $N$ rows ($N=n\times m$), $\bf u$,
$\bf v$, and $\bf V$. Besides, we compact the vectors corresponding
to both field components in a unique field vector, by concatenating
one after another: $\mathbf{q}=({\bf u}^T|{\bf v}^T)^T$, with $2N$
components $q_k$. In that way, the algebraic nonlinear system can be
written as $\mathbf{A}\mathbf{q}=0$, where $\mathbf{A}$ is the
$2N\times 2N$ matrix which depends on the unknown vector through the
nonlinear terms, and we denote as $\mathbf{A}[\mathbf{q}]$, so
that the system of equations takes the form,
\begin{equation}
\label{nonlinear_algebraic_system} \mathbf{A}[\mathbf{q}]\mathbf{q}=0,
\end{equation}
being the rows of the matrix product
\begin{equation}
\label{equations_of_the_nonlinear_system}
E_k=\Sigma_j(\mathbf{A}[\mathbf{q}])_{kj} q_j,
\end{equation}
so that the system is written as $E_k=0,\quad k=1,2,\ldots, 2N$. The
matrix $\mathbf{A}$, even being huge in size, is in practice very
sparse, and it differs from zero at the main diagonal, two
diagonals next to the main one, and two more at the distance $n$
from the main one (this four diagonals appear due to the coupling
terms in the derivatives of the Laplace operator), and also two
more diagonals at a distance $N$ from the main one, due to the
coupling between both field components.

The nonlinear system of equations~(\ref{nonlinear_algebraic_system})
can be solved using the standard globally convergent Newton
method~\cite{Dennis1983,Press}, which builds the solution
iteratively from an initial guess $\mathbf{q}^0$ in the form
$\mathbf{q}^{l}=\mathbf{q}^{l-1}+\delta \mathbf{q},\quad
l=1,2,\ldots$, where the calculation of the so-called Newton step
$\delta \mathbf{q}$ at each iteration involves the solution of the
linear system:
\begin{equation}
\label{Newton_step} \mathbf{J}(\delta\mathbf{q})=-\mathbf{E},
\end{equation}
where $\mathbf{E}$ is the vector obtained by substituting the last
iterate into Eq.~(\ref{equations_of_the_nonlinear_system}), and
$\mathbf{J}$ is the Jacobian matrix defined as $J_{ij}=\partial
E_i/\partial q_j$, and also evaluated substituting the last known
iterate. The Jacobian matrix presents a similar sparse structure
as the matrix $\mathbf{A}$, and it can be calculated analytically.
Obviously, due to a huge size of the matrix $\mathbf{J}$, the system
(\ref{Newton_step}) can only be solved iteratively. Taking into
account that for our particular problem the matrix $\mathbf{J}$ is
symmetric, though in general indefinite, the SYMMLQ
method~\cite{Paige1975} proved to be successful.

Some improvements would be possible in the method, taking the
advantage of the system symmetries. In fact, due to the hexagonal
lattice of holes, the field should be invariant under the rotation
by the angles $l(\pi/3)$, where $l$ is an integer. It would make
possible to solve the problem only in a circular sector of the
amplitude $\pi/3$, imposing periodic boundary conditions at the
borders and homogeneous in the radial direction. The number of points
could be reduced in that case by the factor of six. Another approach,
that also takes an advantage of the lattice periodicity, was developed
by Ferrando {\em et al.}~\cite{Ferrando2003}.

\section{Stationary solutions}

The presence of the external linear potential given by the central
defect and the lattice of air-holes makes the system non-scalable
and its radial symmetry broken. Therefore, the study has to be
carried out by numerical methods. We solve
Eq.~(\ref{model_stationary}) numerically for both scalar (when one
of the components vanishes, i.e. $v=0$) and vector (or
two-component) spatial solitons and obtain the stationary states
of the nonlinear system.

\subsection{Scalar solitons}

For the scalar case, we assume that one of the components is absent
(e.g., $v=0$) and we study a single nonlinear equation of the
nonlinear eigenvalue problem (\ref{model_stationary}). We find a
family of the spatially localized modes--the so-called PCF spatial
solitons--as a function of the mode propagation number $\beta$.
These results are similar to those earlier reported by Ferrando {\em
et al.}~\cite{Ferrando2003}, and the solution can be envisaged as
the fundamental mode of the effective fiber generated by the
combined effect of the PCF refractive index and the nonlinear index
induced by the solution amplitude itself.
\begin{figure}
\centerline{\includegraphics[width=3.0 in]{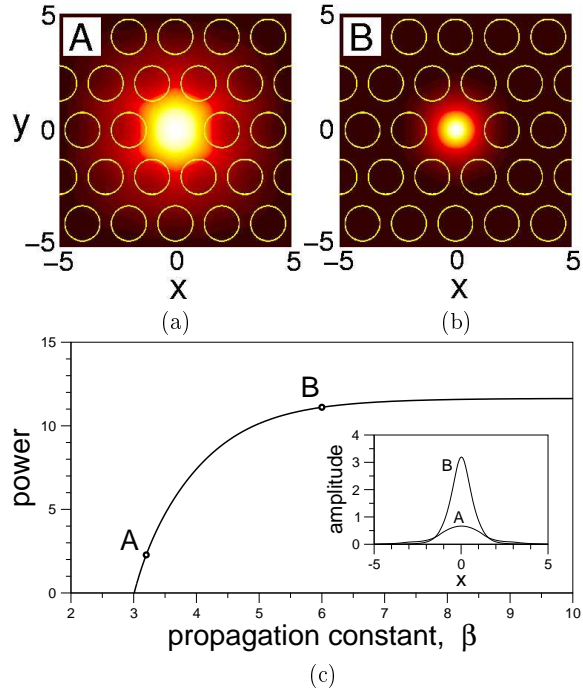}}
\caption{\label{scalar} (a,b) Examples of stationary solutions for
scalar PCF spatial solitons for $\beta = 3.2$ and $\beta = 6$, and
(c) power dependence of the soliton family.  Points A and B in the
power diagram (c) correspond to the examples (a,b), respectively. 
Inset: the corresponding one-dimensional profiles at $y=0$ for both 
examples. 
}
\end{figure}

\begin{figure}
\centerline{\includegraphics[width=3.0 in]{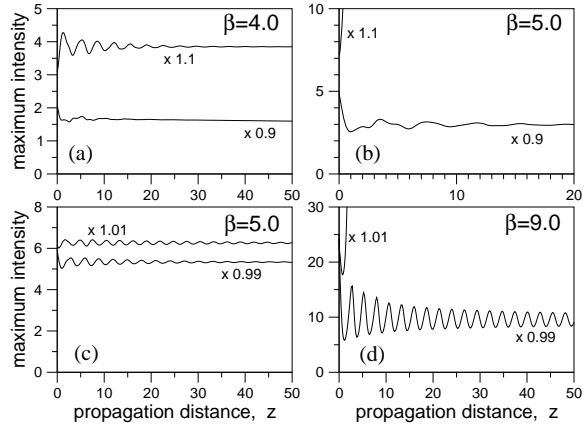}} \caption{
\label{scalar_dynamics} Results of numerical simulations of the
soliton dynamics in the scalar case. The initial stationary solution
is perturbed by two amplitude scalings (indicated in the graphs):
one is higher and the other one is lower than the unity. Shown is
the maximum soliton intensity vs. the propagation distance. }
\end{figure}

Figures~\ref{scalar}(a,b) show two examples of stationary,
spatially localized solutions of the nonlinear model
(\ref{model_stationary}) at $v=0$, which describe scalar spatial
optical solitons as nonlinear modes of PCF. The whole family of
such one-parameter solutions can be characterized by the power $P=
\int u^2 dxdy$, that is plotted in Fig.~\ref{scalar}(c), where the
points A and B correspond to the examples (a,b), respectively. The
material parameters for the PCF are taken as $n_a=1$, $n_s=4$ and
$r=0.75$.

First, we notice that these stationary solutions for scalar
spatial solitons in PCF have been found earlier by  Ferrando {\em
et al.}~\cite{Ferrando2003}, who also mentioned, without a proof,
that such nonlinear modes are stabilized by the lattices of PCF
holes. Indeed, it is well known that in the nonlinear focusing
Kerr media without a nonlinearity saturation, the self-trapped
optical beams are always unstable~\cite{book}. This instability
can manifest itself as the beam spreading, when the input power is
lower than that of the soliton, or the beam collapse, when the
power is larger than the soliton power. As has been mentioned
earlier by Ferrando {\em et al.}~\cite{Ferrando2003}, such a
soliton instability can be suppressed by the presence of the
lattice of holes, because the external potential stops the beam
spreading,  as it happens in a conventional optical fiber, leading
to the existence of a family of stable stationary beams.

In order to demonstrate this feature, we follow the standard
analysis of the soliton stability~\cite{book} and plot in
Fig.~\ref{scalar}(c) the soliton power as a function of the soliton
propagation constant. A positive slope of this dependence indicates
the soliton stability, as will be demonstrated below. In
Fig.~\ref{scalar_dynamics} we present some related numerical
simulations of the dynamics of a perturbed scalar soliton. Some of
the stationary states are scaled by factors slightly higher and
lower than the unity respectively, so as to induce an initial
perturbation, and then propagated using a standard
beam-propagation-method algorithm. The result is that the soliton
behaves stably if its power remains below the maximal limiting
power on Fig.~\ref{scalar}(c).

When the scaling factor is taken higher than unity, a stable
propagation is observed for the solitons of low enough power, as
seen in Fig.~\ref{scalar_dynamics}(a). Nevertheless, for higher
values of the power the soliton may collapse if the scaling factor
is too large [Fig.~\ref{scalar_dynamics}(b)], but it remains stable
for a smaller scaling, see Fig.~\ref{scalar_dynamics}(c). Further
increase in the initial power results in collapse of the beam for
any scaling factor (Fig.~\ref{scalar_dynamics}(d)).

When the scaling factor is taken smaller than unity, the lattice of
holes stops the soliton spreading in all cases, so that the soliton
propagates stable, as is illustrated in all cases presented in
Fig.~\ref{scalar_dynamics}.

\subsection{Vector solitons}

\begin{figure}
\centerline{\includegraphics[width=3.0 in]{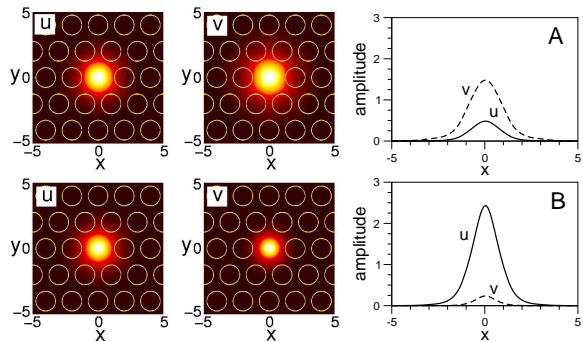}}
\caption{\label{examples} Examples of the two-component stationary
solutions of the model (\ref{model_stationary}) for vector solitons
in PCFs. Two cases correspond to the points A ($\beta = 5, \gamma =
4$) and B ($\beta = 5, \gamma = 8$) in the existence domain shown in
Fig.~\ref{domain} for $\mu = 2$. In each column both components are
shown in the plane $(x,y)$, together with a one-dimensional
$x$-cutoff profile to compare the field amplitudes. }
\end{figure}

\begin{figure}
\centerline{\includegraphics[width=3.2 in]{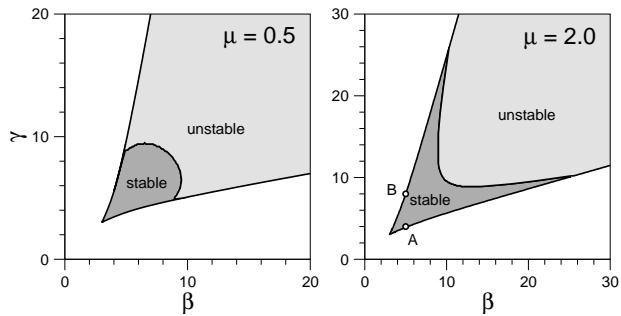}}
\caption{\label{domain} Existence domain for the vector solitons
in PCF in the plane ($\beta$, $\gamma$), shown at two values of
the coupling parameter $\mu$. Labeled points correspond to two
particular examples shown in Fig.~\ref{examples}.}
\end{figure}

Vector solitons in the coupled problem (\ref{model_stationary})
depend on both propagation constants $(\beta,\gamma$) as well as
the material parameters $(\delta,\mu)$. Some examples of the vector
solitons in PCF are presented in Fig.~\ref{examples}, corresponding
to the points A and B marked on the existence domain shown in
Fig.~\ref{domain}. This domain, whose symmetry respect both
parameters $\beta$ and $\gamma$ is evident from the symmetry of both
the equations of the model (\ref{model_stationary}), is plotted in
the plane $(\beta,\gamma)$. The existence domain is limited by two
lines at whose points (bifurcation points) the vector solitons
originate from the scalar solitons; such curves can be regarded as
\emph{bifurcation curves}. When $\mu < 1$, close to the lower
bifurcation curve, the second component decreases becoming a linear
guided mode of the soliton mode in the first (self-guided)
component; the opposite case occurs close to the upper bifurcation
curve where the role of the components is reversed. When $\mu > 1$,
we have the opposed situation with respect to the lower and upper
bifurcation curves. The presence of an effective waveguide
associated with one missing hole in the lattice is the reason that
the propagation constants take a value different from zero, when the
power vanishes; this threshold value corresponds to the eigenvalue of
the linear mode guided by this effective waveguide in the lattice of
air-holes.

Similar to the scalar case, the presence of a periodic lattice of
holes suggests that the vector solitons may become stable in this
system. In this case, the vectorial nature of the system plays an
important role to determine the portion of the domain where the
solutions are stable. As follows from the next section, by applying
the generalized matrix stability criterion, it is possible to
determine the boundary between the stable and unstable regions.
According to that, this boundary is the set of points that fulfill
the marginal stability condition $\det(D)=0$, where $D_{ij}=\partial
P_i/\partial \beta_j$ while $(P_1,P_2)$ and $(\beta_1,\beta_2)$ are
respectively powers and propagation constants for the components
$(u,v)$. In Fig.~\ref{domain} this boundary, as well as both regions
of stability and instability are represented. A number of numerical
simulations were carried out to test the stability of the solutions
in each region. A standard beam propagation algorithm was used and
the stationary solutions of the system were rescaled by a constant
slightly higher that one, so that the peak amplitude of the fields
initially raises over the peak amplitude of the exact stationary
solution. For fields in the stability region, in spite of the
initial growth of amplitude, it becomes stable after certain
propagation distance.

\section{Soliton stability}

Stability of scalar and vector solitons in the coupled NLS equation
can be studied with the matrix stability criterion
\cite{PRE_00,Pel05}. Applications of the matrix criterion depend on
the exact count on the number of eigenvalues of the matrix
Schr\"{o}dinger operators and require careful numerical computations
of a spectral (linearization) problem. Alternatively, a count of the
eigenvalues can be developed in a local neighborhood of the
bifurcation curves, such as the ones shown on Fig.~\ref{domain}.
These computations can be developed analytically, with the
perturbation series expansions \cite{PY00,PY03,PY05}.

\subsection{Scalar solitons}

For simplicity and without loss of generality, we set $n_a=0$ in our
analytical computations. First, we study the stability of scalar
solitons, when $u =\phi(x,y)$ and $v = 0$, where $\phi(x,y)$ is a
solution of the nonlinear eigenvalue problem:
\begin{equation}
\label{scalar-problem} \Delta \phi - \beta \phi + V(x,y) \left(
\delta + \phi^2 \right) \phi = 0,
\end{equation}
We assume that there exists a ground state (positive definite)
solution of the linear problem
$$
\Delta \phi_0 - \beta_0 \phi_0 + \delta V(x,y) \phi_0 = 0.
$$
with the propagation constant $\beta_0$. Applying the local
bifurcation analysis for the nonlinear ground state, we look for
the solutions in the asymptotic form, $\phi = \epsilon [ \phi_0 +
\epsilon^2 \phi_2 + {\rm O}(\epsilon^4)]$ and $\beta = \beta_0 +
\epsilon^2 \beta_2 + {\rm O}(\epsilon^4)$, and obtain the result
\begin{equation}
\beta_2 = \frac{(\phi_0^2, V(x,y) \phi_0^2)}{(\phi_0,\phi_0)} > 0,
\end{equation}
where the inequality follows from the fact that $V(x,y)$ is
non-negative. Therefore, the soliton power (squared $L^2$ norm)
$p(\beta) = \| \phi \|_{L^2}^2 = (\phi,\phi)$ is an increasing
function of the propagation constant $\beta$ near the bifurcation
point $\beta = \beta_0$:
\begin{equation}
\label{power} \frac{d p}{d \beta} = \frac{(\phi_0,\phi_0)}{\beta_2}
+ {\rm O}(\epsilon^2)
> 0.
\end{equation}

Stability for the scalar solitons is determined by the linear
eigenvalue problem, $L_+ u = -\lambda w$ and $L_- w = \lambda u$,
where the linear operators $L_{\pm}$ are defined as,
\begin{eqnarray*}
L_+ & = & \beta - \Delta - V(x,y) \left( \delta + 3 \phi^2(x,y) 
\right) \\
L_- & = & \beta - \Delta - V(x,y) \left( \delta + \phi^2(x,y)
\right).
\end{eqnarray*}
If $\phi(x,y) > 0$ for all $(x,y) \in R^2$, then $L_-$ is
non-negative with the zero eigenvalue $L_- \phi = 0$ due to the
gauge invariance. We shall consider the number of negative
eigenvalues of $L_+$ and apply the earlier results for
one-dimensional solitons~\cite{PRE_00}. It is clear that $L_+$
must have at least one negative eigenvalue, since
\begin{equation}
(\phi,L_+ \phi) = -2 (\phi^2,V(x,y)\phi^2) < 0
\end{equation}
When $\beta = \beta_0$ and $\phi = 0$, the operator $L_+$ has a
simple zero eigenvalue and no negative eigenvalues. Therefore,
according to the perturbation theory, the operator $L_+$ has exactly
{\em one negative eigenvalue} for $\beta > \beta_0$ near the local
bifurcation threshold. The condition for applicability of the
Vakhitov-Kolokolov criterion is satisfied and it suggests stability
of scalar solitons at least near the bifurcation point.

Numerically, we have checked that the number of negative
eigenvalues of $L_+$ does not change and the slope of $dP/d\beta$
is always positive, as shown in the example presented in
Fig.~\ref{scalar}(c). Therefore, the scalar optical solitons in
PCF is stable everywhere for $\beta > \beta_0$.

\subsection{Vector solitons}

Next, we study stability of vector solitons, when $u = \Phi(x,y)$
and $v = \Psi(x,y)$, where $\Phi(x,y)$ and $\Psi(x,y)$ are
real-valued positive solutions of the coupled nonlinear eigenvalue
problem:

\begin{equation}
\label{stationary_equations}
   \begin{array}{l} {\displaystyle
      \beta \Phi =\Delta \Phi + V(x,y) \left( \delta + \Phi^2 + 
		\mu \Psi^2 \right)
\Phi,}\\*[9pt]
        {\displaystyle
     \gamma \Psi =\Delta \Psi + V(x,y) \left( \delta + \mu \Phi^2 + 
		\Psi^2 \right) \Psi,}
   \end{array}
\end{equation}
We consider a local bifurcation of the vector soliton from the
scalar one, and look for solutions in the asymptotic form, $\Phi =
\Phi_0 + \epsilon^2 \Phi_2 + {\rm O}(\epsilon^4)$, $\Psi = \epsilon
( \Psi_0 + \epsilon^2 \Psi_2 + {\rm O}(\epsilon^4))$, and also
expand the eigenvalue, $\gamma = \gamma_0 + \epsilon^2 \gamma_2 +
{\rm O}(\epsilon^4)$, where $\beta$ is an arbitrary parameter, such
that $\beta > \beta_0$. The function $\Phi_0 = \phi(x,y)$ satisfies
the nonlinear eigenvalue problem (\ref{scalar-problem}) for a scalar
soliton. Function $\Psi_0 = \psi(x,y)$ is a ground state solution of
the linear eigenvalue problem:
\begin{equation}
L_0 \psi = \left( \gamma_0 - \Delta - V(x,y) \left( \delta + \mu
\phi^2 \right) \right) \psi = 0,
\end{equation}
where $\gamma_0$ is a function of parameters ($\delta,\mu$) and the
propagation constant $\beta$. The problem for $\Phi_2(x,y)$, $L_+
\Phi_2 = \mu V(x,y) \phi \psi^2$, is always solvable, in the
assumption that the operator $L_+$ for the scalar soliton has one
negative and no zero eigenvalues for any $\beta > \beta_0$. Finally,
from the solvability condition of the linear inhomogeneous problem,
\begin{equation}
L_0 \Psi_2 = 2 \mu V(x,y) \phi \psi \Phi_2 + V(x,y) \psi^3 -
\gamma_2 \psi,
\end{equation}
we derive that
$$
\gamma_2 = \frac{2 \mu (\psi^2, V(x,y) \phi \Phi_2) +
(\psi^2,V(x,y) \psi^2)}{(\psi,\psi)}.
$$
Numerical results show that $\gamma_2 > 0$ for $\mu < 1$ (i.e. the
bifurcation occurs from the lower boundary of the existence domain
on the plane $(\beta,\gamma)$), $\gamma_2 < 0$ for $\mu  > 1$
(i.e. bifurcation occurs from the upper boundary of the existence
domain), and $\gamma_2 = 0$ for $\mu = 1$ (i.e. the existence
domain shrinks on the diagonal $\gamma = \beta > \beta_0$ and
$\gamma_0 = \beta_0$.)

We compute the Hessian matrix of derivatives of individual powers
$P = (\Phi,\Phi)$ and $Q = (\Psi,\Psi)$ with respect to parameters
$\beta$ and $\gamma$. Let $p = (\phi,\phi)$ and assume that
$p'(\beta) > 0$ for scalar soliton with $\beta > \beta_0$. Near
the local bifurcation threshold at $\gamma = \gamma_0$, we have:
\begin{equation}
\frac{\partial P}{\partial \gamma} = \frac{2
(\phi,\Phi_2)}{\gamma_2} + {\rm O}(\epsilon^2) = \frac{\partial
Q}{\partial \beta}, \qquad \frac{\partial Q}{\partial \gamma} =
\frac{(\psi,\psi)}{\gamma_2} + {\rm O}(\epsilon^2),
\end{equation}
such that the determinant of the Hessian matrix is
\begin{equation}
D(\beta,\gamma) = \frac{(\psi,\psi) p'(\beta)}{\gamma_2} - \frac{4
(\phi,\Phi_2)^2}{\gamma_2^2} + {\rm O}(\epsilon^2).
\end{equation}
When $\mu < 1$, we have $\gamma_2 > 0$ and the determinant may
change the sign. Numerical results show that $D > 0$ for $\beta_0
< \beta < \beta_*$ and $D < 0$ for $\beta > \beta_*$ near the
local bifurcation boundary $\gamma = \gamma_0$.

When $\mu > 1$, we have $\gamma_2 < 0$ and the determinant is
always negative near the local bifurcation boundary $\gamma =
\gamma_0$. When $\mu = 1$, we have $\beta = \gamma$ and $P = Q$,
such that $D = 0$.

With the standard linearization, the stability problem for vector
solitons reduces to the matrix eigenvalue problem $\hat{L}_+ {\bf
u} = - \lambda {\bf w}$ and $\hat{L}_- {\bf w} = \lambda {\bf u}$,
where ${\bf u}$ is a two-vector of real parts of the perturbation
and ${\bf w}$ is a two-vector of imaginary parts of the
perturbation, for a real eigenvalue $\lambda$. The matrix
Schrodinger operators are
$$
\hat{L}_+ = \left( \begin{array}{ccc} \beta - \Delta - V (\delta + 
3 \Phi^2 + \mu \Psi^2) & - 2 \mu V \Phi \Psi \\
- 2 \mu V \Phi \Psi & \gamma - \Delta - V ( \delta + 3 \Psi^2 + \mu
\Phi^2) \end{array} \right)
$$
$$
\hat{L}_- = \left( \begin{array}{ccc}  \beta - \Delta - V ( \delta +
\Phi^2 + \mu \Psi^2) & 0 \\ 0 & \gamma - \Delta - V ( \delta +
\Psi^2 + \mu \Phi^2) \end{array} \right)
$$
Since $\hat{L}_-$ is a diagonal composition of two scalar
Schr\"{o}dinger operators, each has a simple zero eigenvalue with
the ground state $\Phi$ and $\Psi$, therefore, the operator
$\hat{L}_-$ is non-negative. Therefore, stability of fundamental
vector solitons is determined by the number of negative eigenvalues
of the matrix operator $\hat{L}_+$, similar to Ref.~\cite{PRE_00}.

We compute the number of negative eigenvalues of the operator
$\hat{L}_+$ near the local bifurcation point. When $\gamma =
\gamma_0$ and $\Psi = 0$, we have
\begin{equation}
\hat{L}_+ = \left( \begin{array}{ccc} L_+ & 0 \\
0 & L_0 \end{array} \right)
\end{equation}
such that the operator $L_+$ has exactly one negative eigenvalue (by
the assumption that the scalar soliton is stable for $\beta >
\beta_0$) and the operator $L_0$ has a simple zero eigenvalue with
the eigenfunction $\psi$. We study bifurcation of the simple zero
eigenvalue of $L_0$ for $\gamma \neq \gamma_0$. Using the same small
parameter $\epsilon$ as in the local bifurcation analysis, we are
looking for solution of the eigenvalue problem $\hat{L}_+ {\bf u} =
\lambda {\bf u}$ by the regular perturbation theory: $u_1 = \epsilon
U_1 + {\rm O}(\epsilon^3)$, $u_2 = \psi + \epsilon^2 U_2 + {\rm
O}(\epsilon^4)$, and $\lambda = \epsilon^2 \lambda_2 + {\rm
O}(\epsilon^4)$.

By algorithmic computations of the regular perturbation theory, we
have the linear inhomogeneous problem for the first-order
correction, $L_+ U_1 = 2 \mu V(x,y) \phi \psi^2$, which is solvable
with the solution $U_1 = 2 \Phi_2$. Furthermore, we have the linear
inhomogeneous problem for $U_2$,
$$
L_0 U_2 = (\lambda_2 - \gamma_2) \psi + 2 \mu V(x,y) \phi \psi
\Phi_2 + 3 V(x,y) \psi^3 + 2 \mu V(x,y) \phi \psi U_1,
$$
with the solvability condition:
$$
\lambda_2 = \gamma_2 - \frac{6 \mu (\psi^2, V(x,y) \phi \Phi_2) + 3
(\psi^2,V(x,y) \psi^2)}{(\psi,\psi)} = -2 \gamma_2.
$$

When $\mu < 1$, we have $\gamma_2 > 0$, such that the zero
eigenvalue of $L_0$ becomes a negative eigenvalue of $\hat{L}_+$.
As a result, we have two negative eigenvalues of $\hat{L}_+$ near
the local bifurcation boundary. Since $p'(\beta) > 0$, we have two
positive eigenvalues of the Hessian matrix when $D > 0$ and one
positive eigenvalue when $D < 0$. In the former case, the vector
soliton is stable, while it is unstable in the latter case.
Therefore, the boundary of $D = 0$ separate the domains of
stability and instability of vector solitons on the plane
$(\beta,\gamma)$ in the assumption that the number of negative
eigenvalues of $\hat{L}_+$ remains unchanged in the entire
existence domain.

When $\mu > 1$, we have $\gamma_2 < 0$, such that the zero
eigenvalue of $L_0$ becomes a positive eigenvalue of $\hat{L}_+$. As
a result, we have only one negative eigenvalue of $\hat{L}_+$ near
the local bifurcation boundary. In the same region, we have exactly
one positive eigenvalue of the Hessian matrix, since $D < 0$.
Therefore, the vector soliton is stable near the local bifurcation
boundary. Numerics show that there exists a curve $D = 0$ in the
existence domain (see Fig.~\ref{domain}(b)), where the positive 
eigenvalue of the Hessian
matrix crosses zero and becomes negative eigenvalue. These curve
approaches the bifurcation curves asymptotically for large
$(\beta,\gamma)$, since $D < 0$ on the bifurcation curves. In the
assumption that the number of negative eigenvalues of $\hat{L}_+$
remains unchanged in the entire existence domain, the curve $D = 0$
separates the stability and instability domains.

When $\mu = 1$, we have $\gamma = \beta$ and the zero eigenvalue of
$L_0$ is preserved as the zero eigenvalue of $\hat{L}_+$ in the
entire existence domain $\beta = \gamma > \beta_0$. This additional
eigenvalue is related to an arbitrary polarization of the vector
soliton in the case $\mu = 1$: $\Phi = \cos\theta \; \phi$ and $\Psi
= \sin \theta \; \phi$, where $\phi$ solves the scalar problem
(\ref{scalar-problem}). The operator $\hat{L}_+$ always has a single
negative eigenvalue (since we have verified numerically that $L_+$
has a single negative eigenvalue for $\beta > \beta_0$). Therefore,
the vector soliton must be linearly stable in the case $\mu = 1$ for
any $\beta > \beta_0$ (excluding the limit $\beta \to \infty$).

\section{Conclusions}

We have demonstrated that stable two-dimensional vector solitons can
be supported by a nonlinear PCF structure with the Kerr
nonlinearity. They constitute a class of two-component spatially
localized modes that bifurcate from their one-component scalar
counterparts and are described by two independent parameters. Both
scalar and vector solitons provide a generalization of the guided
mode trapped in the PCF core to the nonlinear case, being confined
by both linear and self-induced nonlinear refractive indices. The
periodic PCF environment provides also an effective stabilization
mechanism for these localized modes, in a sharp contrast with an
entirely homogeneous nonlinear Kerr medium where both scalar and
vector spatial solitons are unstable and may undergo the collapse
instability. We have applied the analytical matrix criterion for
stability of these PCF vector solitons, and have verified that this
criterion is confirmed by the direct simulations of the soliton
dynamics.

\section{Acknowledgments}

This work has been supported in part by the Australian Research
Council. JRS acknowledges a visiting fellowship granted by the
Direcci\'on Xeral de Investigaci\'on e Desenvolvemento of Xunta de
Galicia (Spain). Both JRS and DEP thank Nonlinear Physics Center
at the Australian National University for a warm hospitality
during their stay in Canberra.

\end{document}